# The Triple Helix Perspective of Innovation Systems

Loet Leydesdorff [1] and Girma Zawdie [2] [•]

**Abstract**

*Alongside the neo-institutional model of networked relations among universities, industries, and governments, the Triple Helix can be provided with a neo-evolutionary interpretation as three selection environments operating upon one another: markets, organizations, and technological opportunities. How are technological innovation systems different from national ones? The three selection environments fulfill social functions: wealth creation, organization control, and organized knowledge production. The main carriers of this system—industry, government, and academia—provide the variation both recursively and by interacting among them under the pressure of competition. Empirical case studies enable us to understand how these evolutionary mechanisms can be expected to operate in historical instance. The model is needed for distinguishing, for example, between trajectories and regimes.*

---

[1] Amsterdam School of Communications Research (ASCoR), University of Amsterdam, Kloveniersburgwal 48, 1012 CX Amsterdam, The Netherlands; loet@leydesdorff.net ; http://www.leydesdorff.net .
[2] Girma Zawdie, Department of Civil Engineering, University of Strathclyde, John Anderson Building, 107 Rottenrow, GLASGOW, G4 ONG; g.zawdie@strath.ac.uk
[•] Corresponding author

**Introduction**

Far from being a program running parallel to or competing with the national innovation system (NIS), the Triple Helix of university-industry-government relations was introduced to bring out the depth and complexity of the innovation process as a recursive interaction system underlying the knowledge-based economy, and thus to enhance the exploration and exploitation of this knowledge base on conceptual and empirical grounds (Abramowitz & David, 1996; David & Foray, 1995 and 2002). While NIS is ultimately an institutional program focused on wealth creation at the national—or *mutatis mutandis*, regional—level, Triple Helix provides a model of the structure and dynamics underlying the innovation system functioning at various levels. Unlike NIS (or RIS), the Triple Helix model does not presume a geographically delineated system, but it provides a framework for investigating empirical questions at a level of "systemness," defined in terms of regimes and trajectories. Are innovation systems aligned at the level of nations, sectors, regions, etc., and if so, to which extent? Can the variation be explained in terms of underlying structures—that is, selection environments—and their recursive interactions?

This special issue covers contributions selected from the proceedings of the 7$^{th}$ International Conference on the Triple Helix that was hosted by the University of Strathclyde in Glasgow in June 2009. The Conference theme on Triple Helix as a framework for addressing the global agenda of innovation, competitiveness, and sustainable development raises questions, *inter alia*, about the theoretical adequacy and empirical validity of Triple Helix as a policy model. How can the Triple Helix model be extended beyond the trilateral relationship between university, industry and government to explain the complex features of the dynamics in the innovation system? In what ways has the introduction of the Triple Helix concept (Etzkowitz & Leydesdorff, 1995, 2000) added to the ongoing efforts of explaining innovation as a systemic phenomenon?

A network analysis cannot inform us about the dynamics, but only about evolving institutional relations. In our opinion, it is rather the "forces of motion" underlying the institutional linkages—namely the way knowledge infrastructures evolve and can have implications for technological opportunities; and the way the political economy evolves to determine the institutional mechanisms for the selection of technological trajectories—that bear out Triple Helix as a dynamic system of three selection mechanisms operating upon one another. These selection mechanisms—markets, organizations, and technological opportunities—change characteristically over time as do the cultural and behavioural patterns of the actors engaged in the interactions involving the process of knowledge generation and knowledge sharing and exchange at the level of each system. The Triple Helix perspective enables us to study innovation systems in empirical terms (to which extent can which arrangement be considered as a system?) and analyze best practice trends (are more knowledge-based arrangements possible?).

**The evolution of systems thinking in innovation theory**

How can the Triple Helix concept be related to the programme of studying "national innovation systems"? How are the concepts of innovation systems in these two research programmes different and yet related? Lundvall (1988) first introduced the concept of "national systems of innovation" by elaborating on Christopher Freeman's (1987) study entitled "*Technology, policy, and economic performance: lessons from Japan*," in which the latter had argued that Western nations could learn from Japan's experience in the coordination, at the national level, of S&T policies orchestrated by the Japanese Ministry of Trade and Industry (MITI) in the preceding decades. Yamauchi (1986) characterized the Japanese experience as a textbook-model: partners in the Japanese system knew what was expected technologically in order to meet (economic) demands and (political) objectives. In this integrative model, university-industry-government relations were synchronized *ex ante*, however impicitly, at the national level.

In Western countries, this *ex ante* synchronization at the national level had been lost because of an ongoing internationalization during the post-War priod. Internationalization of the American corporations during the 1950s was followed by a wave of mergers among European companies during the 1960s. Knowledge-intensive industries were particularly prone to internationalization. The "Sputnik shock" of 1957 had made policy makers aware that S&T policies needed international coordination (York, 1970). The OECD, which was originally initiated for the distribution of Marshall aid after WW II, was transformed into a civil center of coordination—alongside NATO—among national S&T policies.

Most Western European countries developed S&T policies during the 1960s. The global oil-crises of the 1970s enhanced the tendency towards the internationalization of S&T policies. In the 1980s, the EU increasingly began to play a role in innovation policies on the initiative of European presidents such as Jacques Delors, who favored unification on the basis of the cultural heritage of Europe in terms of S&T, both for ideological and economic reasons.

Evolutionary theorizing about innovation emerged alongside these policy initiatives during the late 1970s. Nelson & Winter's article entitled "In search of a useful theory of innovation" (1977) can be considered seminal to the shaping of evolutionary economics (cf. Nelson & Winter, 1982). The theoretical model was soon followed by Pavitt's (1984) empirical studies on innovation patterns using innovation statistics and Freeman's (1982) contributions at the level of the OECD. With the exception of Sahal (1981, 1985), however, these scholars tended to avoid invoking "systems theory," apparently because of the non-empirical—one might even say anti-empirical—inclinations dominant in this research tradition. Although innovation was considered to be systemic, the focus on empirical studies and, therefore, the need for developing measurable indicators was considered a primary objective. Andersen (1994) noted that even the question of "what is evolving" was not yet properly answered by Nelson and Winter (1982). The focus remained on entrepreneurship; and the dominant theory accordingly became a theory of the firm as the carrier of the innovation process (Casson, 1997).

When Lundvall (1988) proposed that the nation be considered as the first candidate for the study of innovation systems, he formulated this claim carefully in terms of a heuristics:

> The interdependency between production and innovation goes both ways. […] This interdependency between production and innovation makes it legitimate to take the

> national system of production as a starting point when defining a system of innovation (Lundvall, 1988, p. 362).

The idea of integrating innovation into production at the *national* level has the advantage of providing the analyst with an institutionally demarcated system of reference. If the market is continuously upset by innovation, can the nation then perhaps be considered as another, albeit institutionally organized (quasi-)equilibrium (Aoki, 2001)? Lundvall, furthermore, proposed to consider the *interactions* between two selection mechanisms in user-producer relations (market dynamics and transaction costs) as a different micro-economic foundation of theorizing (instead of the conventional agent-based economic profit maximization).

The specification of the nation as a well-defined system of reference enables evolutionary economists to study at the macro level, for example, the so-called "differential productivity growth puzzle" which is generated by the different speeds of development among the various industrial sectors (Nelson and Winter, 1975). The problem of the relative rates of innovation cannot be defined properly without the specification of a system of reference that integrates different sectors of an economy (Nelson, 1982, 1994). The solutions to this "puzzle" of differentiation can accordingly be expected to differ among nation-states (Lundvall, 1992; Nelson, 1993).

**In search of a definition of innovation systems**

Although the emergence of transnational levels of government, like in the European Union, together with an increased awareness of regional differences within and across nations, have changed the functions of national governments (Braczyk *et al*., 1998), integration at the national level continues to play a major role in systems of innovation (Skolnikoff, 1993). The historical progression towards internationalization varies across countries.

For example, after its opening to the "market system" in the early 1990s, China further developed its policies deliberately in terms of a national system of innovations, while the European Commission became disenchanted with this concept increasingly in the early 90s: a "knowledge-based economy" develops at the supra-national level. Reducing the concepts to the geographical level of nations or regions may, therefore, fail to address the new (knowledge-based) dynamics which tends to transcend geographical boundaries.

In a workshop on the subject of the knowledge-based economy in 1994, Abramowitz and David (1996: 35) suggested that *codified* knowledge should be made central to the analysis of the modern economy, and formulated as follows:

> Perhaps this single most salient characteristic of recent economic growth has been the secularly rising reliance upon *codified* knowledge as a basis for the organization and conduct of economic activities, including among the latter the purposive extension of the economically relevant knowledge base. While tacit knowledge continues to play critical roles, affecting individual and organizational competencies and the localization of scientific and technological advances, codification has been both the motive force and the favoured form taken by the expansion of the knowledge base.

Analytically, this focus on *codified* knowledge demarcated the new research programme of innovation systems from the older concept of a "knowledge economy" with its focus on knowledge workers and hence embodied knowledge (Cooke, 2002; Machlup, 1962; Penrose, 1959). Embodied and tacit knowledge is embedded in contexts (Bowker, 2005; Collins, 1974; Polanyi, 1961; Zuboff, 1988), while codified knowledge can be decontextualized, and therefore, among other things, traded on a market (Dasgupta & David, 1984). The metaphor of a knowledge-based economy appreciates the increased importance of organized R&D in shaping systems of innovation. The knowledge production function has become a *structural* characteristic—and, therefore, a relevant selection environment—of the modern economy (Schumpeter, 1939, 1943).

Whereas a knowledge-based economy develops as a dynamic system at the global level, thus transcending national or geographical boundaries, wealth from knowledge has to be retained locally. National and regional systems of innovation can thus be considered as retention mechanisms of a self-organizing system that develops at the global level. Government policies hence can no longer be efficient when developed exclusively in terms of economic parameters; the economic environment is no longer the only relevant one. Technological opportunities provide an additional selection environment. For example, as Barack Obama formulated in one of his campaign speeches:

> "[T]his long-term agenda […] will require us first and foremost to train and educate our workforce with the skills necessary to compete in a knowledge-based economy. We'll also need to place a greater emphasis on areas like science and technology that will define the workforce of the 21$^{st}$ century, and invest in the research and innovation necessary to create the jobs and industries of the future right here in America."[3]

At the micro-level one is well aware of this trade-off between different selection mechanisms when one uses price/performance relations instead of the price criterion when buying technological devices such as personal computers: one should not buy the cheapest ones! A relevant question for government policies, however, follows: can wealth be retained in geographical systems, or are national systems of regulation and legislation mainly redistributors of wealth which is generated in other systems? Should a region such as Piedmont, for example, be considered as the frame of reference because it was administratively so defined as a region (OECD, 2009) or should one optimize possible synergies across regional borders between, for example, Piedmont and Lombardy? Previously taken-for-granted systems of reference are no longer given, but can be reconsidered in the light of technological opportunities, patent portfolios, knowledge infrastructures, etc.

Upon studying the emergence of biotechnology during the 1980s, Carlsson and Stankiewicz (1991) proposed that technologies shape their own innovation systems as contexts ("selection environments") needed for the further development of technological innovations as a system. How are technological innovation systems different from national ones? The question of whether systems of innovation are technologically or geographically integrated is pertinent because the structures of a system determine the causalities prevailing in it. Are governments able to "steer" technological developments? Can government incentives be considered as independent variables or are they mainly feedback on systems which are driven by other potentially non-linear dynamics? Thus, the question of technological determinism can be reformulated as an empirical one: to what extent do various agents and mechanisms control the ongoing developments?

---

[3] Remarks of Senator Barack Obama: "Change That Works for You," June 9, 2008, at http://www.barackobama.com/2008/06/09/remarks_of_senator_barack_obam_76.php .

The dynamics of an innovation system are non-linear because they are based on interactions between (economic) demand, (political) objectives, and (technological) opportunities, and also because of path-dependencies in all systems of reference. National patterns (strengths and weaknesses, natural endowments, etc.), for example, shaped systems at national levels between approximately 1870 and 1970. Thus, while car manufacturing is important in Germany's industrial structure, it hardly plays a role in a neighbouring country such as the Netherlands with its service-oriented economy.

What determines a system? A system results from the ways in which selection mechanisms operate given the wide range of possibilities arising from the Schumpeterian phenomenon of "creative destruction." In a knowledge-based economy, three selection mechanisms are continuously recombined, generating successive levels of innovation and technological trajectories: the economic mechanism of the market; the political mechanism of control over resources; and the mechanism for the generation of new knowledge as potentially innovative.

Selection (unlike variation) is a deterministic operation. Variation can be more or less random. Selection exhibits an instance of the selection mechanism operating upon the variation because of a (hypothesized since latent) structure in a system operating. Over time, recursive selections can shape technological trajectories (Nelson & Winter, 1982). Technological trajectories can be considered as the operations of selections upon previous selections; some selections can then be stabilized along a trajectory. Some stabilizations of trajectories along specific pathways can further be selected for globalization into a technological regime.

Thus, three evolutionary mechanisms can be specified along the time axis: selection (mechanism of the market); stabilization (mechanism of control and regulation); and globalization (mechanism of knowledge generation and exchange). These selection mechanisms operate upon the variations in different dimensions from the perspective of hindsight: given a variation, selection can operate. The resulting system is constructed bottom-up, but control operates increasingly top-down as a system is further shaped.

For example, the national systems as they emerged in Europe, the USA, and Japan in the second half of the 19$^{th}$ century,[4] stabilized the workings of markets within national systems by border controls, legislation and regulation, national systems of intellectual property protection, etc. Two selection mechanisms—markets and national politics—could shape national systems or, in other words, political economies. At the global level, these national systems developed along trajectories which allowed for competition, for example, as colonial powers. This system of nations, however, was thoroughly upset by the competition among global regimes (communism, liberal democracy, and fascism) during WW II and the Cold War thereafter.

The fading away of these conflicts set a third mechanism—organized knowledge production and control—free as a globalizing coordination mechanism and relevant selection environment, that is, as a structural component no longer to be considered as exogenous. Along each two of these selection mechanisms, trajectories can be formed. However, the operation of the third mechanism may continuously disturb a co-evolution between the other two. Let us specify the possible co-evolutions or mutual shapings between two of the three functions: knowledge exploration, knowledge exploitation, and organizational control. A knowledge-based economy can be considered as based on interactions between the two (traditional) drivers of a political economy with the new dynamics of knowledge production, diffusion, and control (Lengyel & Leydesdorff, 2010). There are three major aspects to these interactions.

First, there is the synergy between economic knowledge exploitation and organizational control at, for example, the national level or any other geographically delineated systems level (e.g., a region) that can be expected to contain a political economy. Second, a synergy between (economic) knowledge exploitation and knowledge exploration may shape a technological (!) trajectory. Since the latter may evolve across national boundaries, transnational corporations, for example, can serve as the carriers. Third, a co-evolution between knowledge exploration and organizational control can lock an innovation system into a national system of procurement like in the case of medical technologies or (traditional) energy systems. The former Soviet-Union provided an example par excellence in which the mechanism of the market was relatively lamed, whereupon the state apparatus emerged between (academic) knowledge exploration and knowledge exploitation within strictly defined and maintained geographical boundaries.

---

[4] The unification of Germany and Italy in 1870 completed a system of nation states which had been shaped in the period before, for example, during the American civil war (1860-1865) and the Meji restoration in Japan (1863).

The shaping of the regime of a knowledge-based economy assumes that the three selection mechanisms can operate upon one another. The imposition of boundaries by governance can be appreciated as functional specifically to the retention of wealth from knowledge. However, the system operates in such a way that selection mechanisms are no longer institutionally secured (like in a nation state, a firm or a discipline), but are evolutionarily defined in terms of functions: markets operate pervasively, control mechanisms are in place and further developing, and knowledge exploration is socially organized.

In other words, the knowledge-based system tolerates the complexity of potentially divergent and differently rationalized developments. The differences in the rationalities can be considered functional to the further development of the system, but since the knowledge-based system is relatively globalized, it can be expected to remain in flux and can no longer be coordinated *ex ante*. Stabilization of this globalizing system (e.g., into a national system of innovations) presumes the choice of an analytical perspective and therefore mitigation of the prevailing complexity. Such a choice can analytically be heuristic, enabling us to raise empirical questions. However, both the perspectives and the subjects under study contain uncertainty, so that the selection environments are no longer to be considered as given, but as hypotheses which enable us to guide the analysis in a discourse.

Political discourse differs from scientific discourse, while economic exchanges are guided by price incentives. The three coordination mechanisms stand analytically orthogonal, but in innovative practices they interact. These interactions can be considered as instantiations which empirically condition and enable further developments in all three dimensions, but asymmetrically and asynchronously. In other words, a system of innovations can only be specified as a model which integrates discursive models at interfaces; these systems are not hardwired; the reality of such a system remains analytical. "What we see, is not what we get" and depends also on one's perspective. Each perspective allows for an additional set of possibilities and can therefore inform the others; much as in the above example of price/performance as a criterion superior to the price mechanism in the case of knowledge-intensive devices (but not necessarily when buying oranges). Analogously, political discourse faces the huge task to reflexively understand its own priority in relation to ongoing interactions with both economic restructuring and scientific uncertainty. Appreciation of alternative options in the other dimensions, however, opens also the self-reference in each system to innovation.

For example, in a recent study of the Hungarian system of innovations, it was found that the assumption of national integration was no longer fruitful for explaining the differences among regions (Lengyel and Leydesdorff, 2010). Unlike the Netherlands (Leydesdorff *et al*., 2006), the national level in this country no longer added to the "systemness" of innovations in three sub-national systems: the metropolitan system around Budapest, the western part of the country which has increasingly been integrated into Western-European and international systems of innovation, and the eastern part of the country which is largely integrated according to the old (politically controlled) model which predated the transition of the 1990s. The assumption of "systemness" of innovation can thus be considered as an hypothesis.

"Systemness" assumes that synergy is generated in the relations among subsystems. Like the systems, the subsystems should not be reified. Subsystems are functional insofar as they serve the reproduction of the system; functions in composed systems can be specified as subdynamics. Three subdynamics are suggested by the Triple Helix model as crucial: the economic dynamics of the market, the political dynamics of control, and the socio-cognitive dynamics in the production of organized knowledge. Whitley (1984), for example, specified "organized knowledge production and control systems" as disciplinary combinations of the latter two dynamics in a knowledge infrastructure. Schumpeter (1939, 1942) specified the combination of knowledge production and market dynamics as "creative destruction," which provides the basis for changes in technological trajectories. The relations between political and economic dynamics have been the focus of theorizing about political economies in both Marxist and non-Marxist traditions.

How do political economies change under the pressure of technological trajectories? How are technological trajectories upset when new technological regimes emerge? Path dependencies in the evolving systems of innovation are induced by the asynchronicities among the three coordination mechanisms. Two of the three mechanisms may at any time click into a co-evolution and then mutually shape a trajectory. The third mechanism can be expected to provide the dynamics. A lock-in, for example, between a technology and the market can be hyper-stabilized or gradually destabilized by this third dynamic depending of the prevailing sign of the feedback or feedforward (Dolfsma & Leydesdorff, 2009),

Freeman and Perez (1988) developed a model of structural adjustment policies in which cycles are induced periodically by new key factors of the economy. However, this model remained a dialectical model of the development of the political economy under the pressure of otherwise exogenously defined technological developments induced by rapidly falling prices in factor inputs. The production of these new resources at the supra-institutional level by organized knowledge production and control was not yet deconstructed and incorporated into the model.

When the model allows for interaction among the *three* subdynamics of the system—with one of these subdynamics considered as an exogenous variable conditioning the coevolution between the other two—a Triple Helix model with three-way interactions can be hypothesized. The new model (based on a neo-evolutionary interpretation of the Triple Helix in terms of interacting functions) enables us not only to envisage trajectory changes in the downswing phases of the economy, that is, at the end of cycles, but also the induction of regime changes in the technological environment, giving rise to the development of innovative products and processes of strategic significance, as in the case of renewable energy systems.

While carbon-based energy production and consumption has become increasingly a burden to the current production system and its natural environment, one can envisage the increasingly rapid replacement of carbon-based energy-carriers with energy sources which are virtually unlimited such as solar or geothermic energy. Nevertheless, the introduction of these new energy sources requires adaptation of existing markets and political control structures which may take decades to emerge. Thus, the knowledge-based economy system is both extremely buffered and flexible: new developments continuously emerge, but remain under heavy selection pressures. Old regimes do not give way without first attempting to encapsulate hyperselectively new developments by further differentiation (Bruckner *et al.*, 1994). The more entrenched a system is, the more resilient it can be expected to be against regime change.

For example, Douglas introduced the DC3 as a new paradigm for propeller aircrafts in 1936. It took Boeing, its main competitor, only two years to shape a competitive aircraft (the 1938-Boeing 307 "Stratoliner"). When the next model DC4 (1942) set the standards of the new regime (with four propellers), Boeing disinvested in its existing development capacities and produced the 1944-Boeing 377 "Stratocruiser" in accordance with the specifications of the new paradigm, such as a steel-based closed-body aircraft (Frenken & Leydesdorff, 2000, p. 340).

When in a later phase of the development of civil aircraft, Boeing introduced the jet-engine based wide body aircraft (707) in 1957, Douglas could develop a competing aircraft (the DC8 in 1958 as a follow-up of the DC7 of 1953) in a few years time, but the propeller aircraft remained the main competitor at the paradigmatic level. Only when Airbus in 1981 accepted the textbook of the Boeing model, a new paradigm for aircraft development became increasingly dominant. Thus, it took more than 20 years to perform the transition; an entrenched paradigm exhibits resilience against change.

Note that control at the regime level is no longer exerted at the level of individual corporations (such as Boeing) or individual nations, but by interacting dynamics at the next-order level of global developments of technologies and markets. Nation states with national aircraft industries may or may not be successful in retaining wealth from these developments. For example, Fokker in the Netherlands failed to make the necessary transition and remained hesitant about the paradigm choice after the transition had been made. It continued to develop both the propeller-based F50 and the jet-engine based F100 and thus overstretched its capacity or that of the Dutch government and knowledge infrastructure.

Increasingly, governments can be reflexive on their positions in the complex dynamics in which *ex ante* synchronization is no longer expected. As the evolutionary perspective prevails, a richer model in the form of the Triple Helix can be entertained, in which the specific position of a nation can be assessed in relation to other possible positions in a distribution of governmental efforts; for example, at the supra-national level (Laredo, 2003). In other words, the selection environments can be considered as distributions which can be assessed in terms of the uncertainties that they are expected to contain.

**Empirical studies and simulations using the Triple Helix model**
We have argued that the Triple Helix perspective can be elaborated into a neo-evolutionary model which enables us to recombine sociological notions of meaning processing in different discourses, economic theorizing about exchange relations, and insights from science and technology studies regarding the organization and control of knowledge production. Communicative competencies developed and appreciated at the supra-individual level can be expected to determine and constrain the innovative capacities of knowledge-based systems. Can the differently codified communications be translated into each other and can these translations be appreciated mutually?

Where the translations can resonate with the historically embedded (recursive) communication structures, the further codification of meaning in scientific knowledge production can be expected to add value to the economic exchange relations (Foray, 2004; Frenken, 2005). Triple Helix serves us mainly as a heuristic model for modeling these interactions. Its abstract and analytical character enables us to explain current transitions towards a knowledge-based economy as a new regime of operations. In other words, this neo-evolutionary version of the Triple Helix model operationalizes the general notion of a knowledge-based economy as a self-organizing system (Krugman, 1996) in terms of three relevant selection environments.

The differentiation in terms of selection mechanisms can be both horizontal and vertical. Vertically, the flux of communications is constrained by the institutional arrangements that have been shaped in terms of stabilizations of previous communicative structures. Horizontally, the coordination mechanisms can be of a different nature because they can be expected to use different codes. For example, market transactions are different from scientific communications. Market transactions can also be cross-tabulated with organizational (control-) hierarchies (Williamson, 1975; Lundvall, 1988). While the control mechanisms at interfaces can be considered as functional for the differentiation among communications, the hierarchy in the organization may help reduce the problem of coordination between functions to a multi-level problem within the institutional dimension.

In summary, the functional perspective is different from the institutional one. Functional communications evolve; institutional relations function as retention mechanisms which respond to functional incentives. However, the functions are not given, but have to be specified. Their epistemological status remains that of (more or less informed) hypotheses. Thus, one can study a Triple Helix system at different levels and from different perspectives. For example, one can study university-industry-government relations from a (neo-)institutional perspective (e.g., De Rosa Pires and De Castro, 1997; Etzkowitz *et al*., 2000; Gunasekara, 2006) or one can focus on the relations between university science and the economy in terms of communications (e.g., Langford *et al*., 1997). Different interpretations of the Triple Helix model can be at odds with each other, but still inform the model. Each metaphor stabilizes a geometrical representation of an otherwise more complex dynamics.

Competing hypotheses derived from different versions of the Triple Helix model can be explored through formal modeling and appreciated through (neo-)institutional analysis. Case studies inform the modeling efforts about contingencies and boundary conditions, while simulation models enable us to relate the various perspectives. Such translations potentially reinforce the research process by raising new questions; for example, by comparing issues across different contexts and/or with reference to emerging phenomena. In the model, the three strands of the Triple Helix are declared as (formally equivalent) selection mechanisms, but they are substantially very different. The selection mechanisms are expected to operate asymmetrically.

The one strand (university) is institutionally less powerful than the other two strands. Furthermore, the other two strands (government and industry) are increasingly and indirectly co-opting the university in a variety of ways. However, the university has specific strengths: it is salient in providing the other two systems with a continuous influx of new discursive knowledge (e.g., publications and patents) and new knowledge carriers (students). From this perspective, the university can be considered as a main carrier of the knowledge-based innovation system (Godin and Gingras, 2000). Knowledge-based fluxes continuously upset and reform the dynamic equilibria sought by the two other strands of the political economy.

**The Triple Helix and empirical studies**

In our opinion, the neo-evolutionary version of the Triple Helix model is sufficiently complex to encompass the different perspectives of participant observers (e.g., case histories) and to guide us heuristically in searching for options newly emerging from the interactions. What is the contribution of this model in terms of providing heuristics to empirical research?

First, the neo-institutional model of arrangements among different stakeholders can be used in case study analysis. Given the new mode of knowledge production, case studies can be enriched by addressing the relevance of the third major dimension of the model. This does not mean to disclaim the legitimacy of studying, for example, bi-lateral academic-industry relations or government-university policies, but one can expect more interesting results by observing reflexively the interactions among the three subdynamics. In other words, one can increase the relevance of a study by reflecting on how the third context may add to the richness of the conclusions.

Secondly, the model can be informed by the increasing understanding of complex dynamics and simulation studies from evolutionary economics (e.g., Malerba *et al.*, 1999; Windrum, 1999). Thirdly, the Triple Helix model adds to the meta-biological models of evolutionary economics, the sociological notion of meaning being exchanged among the institutional agents (Habermas, 1987; Leydesdorff, 2010; Luhmann, [1984] 1995).

Finally, on the normative side of developing options for innovation policies, the Triple Helix model provides us with an incentive to search for *mismatches* between the institutional dimensions in the arrangements and the (hypothesized!) social functions carried by these arrangements. The frictions between the two layers (knowledge-based expectations and institutional interests), and among the three domains (economy, science, and policy) provide a wealth of opportunities for puzzle solving and innovation.

The evolutionary regimes are expected to remain in transition because they are shaped along historical trajectories. Shifts in a knowledge-based regime can be expected continuously to upset the political economy and the market equilibria as different subdynamics. Conflicts of interest can be deconstructed and reconstructed in terms of these different coordination mechanisms, first analytically and then perhaps also in practice in the search for solutions to problems of economic productivity, wealth retention, and knowledge growth.

**Salient features of this special issue**

The Triple Helix model has been embraced by policy-makers because of its neo-corporatist overtones and its emphasis on collaboration at local, regional, and national levels. Unlike the *a priori* choice for the national level, the Triple Helix model can be appreciated at various levels of geographical integration. The self-organizing model of the Triple Helix, however, does not privilege any perspective *ex ante*. The knowledge dynamics tends to globalize and thus to uncouple from institutional conditions (Leydesdorff & Fritsch, 2006).

Grasping wealth from retaining these dynamics requires a more informed and reflexive approach. Nationally and regionally motivated aspirations may, however, be counterproductive to a knowledge-based dynamics in metropolitan areas (Florida, 2002). For example, the structure of different regions in Spain, such as Catalonia or the Basque country, can be expected to require different innovation policies. Whereas Catalonia claims (sub)national integration at the level of this region for political reasons, the system may be integrated nationally and internationally more than envisaged locally (Riba-Villanova & Leydesdorff, 2001). The Basque country, on the other hand, which is hitherto less integrated in the knowledge dynamics, may find it easier to develop a regional innovation system (Moso and Olazaran, 2002).

Whereas the national and regional innovation systems would seek to retain the benefits of globalizing knowledge within specified geographical boundaries, the Triple Helix model underpins the study of innovation systems at various levels in terms of institutional and functional categories. It can thus be argued that the Triple Helix perspective has enriched the conceptual and empirical dimensions of innovation as a systemic phenomenon, thus potentially improving the effectiveness of innovation policies at regional and national levels, and in a system where knowledge production is being increasingly globalised.

As is apparent from the discussion in the foregoing of this paper, there is a wide range of issues of theoretical and empirical significance arising from the Triple Helix approach to the innovation system. It would, therefore, be overly ambitious for a special issue like this to be comprehensive in its coverage of the major aspects of the Triple Helix perspective of innovation systems. Rather, the aim of the issue is to contribute to the ongoing discussion by shedding light on some key issues, making them more discernible; and teasing out issues of policy relevance that could be promoted within the Triple Helix framework.

The four papers included in this special issue have a common strand running through them—namely, the regional dimension of the Triple Helix innovation system. This common strand set in different contexts, shows the robustness of the Triple Helix model as a heuristic for empirically investigating the complex dynamics underlying the innovation process at regional level. Two of the papers—one by Helen Lawton-Smith and Sharmistha Bagchi-Sen, and the other by Carl-Otto Frykfors and Hakan Jonsson—discuss Triple Helix as a facilitator for the emergence of industrial clusters as a basis for regional development, albeit from different perspectives.

The first study entitled "Triple Helix and Regional Development: a perspective from Oxfordshire in the UK" attributes the concentration of biotechnology activities in the region to the way the three dimensions of the Triple Helix have interacted to determine the distinctiveness of the region in terms of the evolution of political economy, knowledge infrastructure, and technological trajectory. The emergence of dominant factors in this process have made Oxfordshire a favourable location for the development of the biotechnology cluster. While university, industry, and government are all important for the emergence of the biotech cluster in Oxfordshire, Lawton-Smith and Bagchi-Sen found the dominant factor that particularly underpinned the biotechnology capability of the region to be not the role played by Oxford University as a world centre for biomedical research, but the availability of skills and talents in science and technology in the region.

Frykfors and Jonsson discuss in their study entitled "Reframing the Multilevel Triple Helix in a Regional Innovation System: a case of systemic foresight and regimes in renewal of Skåne's food industry" the reconstruction of a relatively low-tech mature industry cluster. The authors highlight the importance of a multi-level approach to Triple

Helix policy to engage in systemic interaction all Triple Helix actors across the value chain, thus paving the way for the emergence of an overarching innovation community in the sector.

A characteristic feature of low-tech mature industrial clusters is the prevalence, at various levels, of sectoral activities of differentiated social and technological cultures that have ossified over the period the sector has evolved with boundaries between activities that constrain interaction and collaboration between Triple Helix actors. In such cases, the development of Triple Helix networks can blur the boundaries between spheres of activities; diminish the asymmetry in the distribution of information and hence the transactions cost of interactions between actors in the sector; and facilitate the emergence of innovation communities and prospects for sustainable development in the sector.

Thus, as Frykfors and Jonsson argue, the transformative effect of Triple Helix policy on regional economies is not exclusive to high-tech clusters as is often presumed, but has also a significant role in the development of regions with low-tech mature industry clusters incorporating heterogeneous activities, as in the case of the food cluster in the Skane region in Sweden. However, where vested interests prevail against cultural transformation, it is important that regional policy is designed in a multilevel Triple Helix framework, so that the challenge of such forces is countervailed at all levels of sectorial activities.

It may be asked as to how application of the Triple Helix policy mechanism to regional industry clusters translates into knowledge-based innovation. In his contribution entitled "Regional Innovation Systems: Development Opportunities from the 'Green Turn'," Philip Cooke takes this issue on board on the basis of the literature of regional innovation system to which he has been and is a principal contributor. In a bid to explore the pathway for "green innovation," he argues that Triple Helix interactions could be integrated into regional innovation systems to exploit the benefits of synergy arising from the "transverality" of inter-cluster knowledge flows. "Green innovation," which would respond to the global challenges of climate change and resource waste and degradation, could thrive where related clusters (as in the heterogeneous set of renewable energy cluster) provide opportunities or "platforms" for cross-fertilization of knowledge.

Central to Cooke's argument of regional innovation and regional development through cross-pollination of technology platforms is the concept of "related variety" deriving from evolutionary economic geography. Given platform industries as integrated regional clusters based on "related variety" or "knowledge proximity" (as in the case of renewable clusters focused, for example, on wind turbines, solar thermal and photovoltaics, greening engineering, etc.), Triple Helix interactions between clusters would prompt the emergence of innovation at cluster interfaces, making the region the locus of specific categories of strategic innovations (as in the case of "green innovations") that are capable of shifting technology paradigms.

The principle underlying Cooke's analysis of opportunities for the emergence of "green innovation" can also be seen, if in part, in the Lawton-Smith and Bagchi-Sen study on the biotechnology cluster in Oxfordshire. In both cases, essential differences in regional capabilities are recognized; but whereas Lawton-Smith and Bagchi-Sen empirically attribute the location of biotechnology cluster in Oxfordshire to the availability of skills and talents in science and technology and also, if secondarily, to the proximity of Oxford University, Cooke invokes the analytical conceptual framework of path dependency to explain the selection of "new combinations," and hence the pace and direction of "green innovation," based on existing capabilities.

On the other hand, the Frykfors and Jonsson study on food cluster in the Skane region may have an interestingly unconventional implication for the path dependency argument, as its findings suggest that the pace and direction of innovation would be prompted not by existing conditions and capabilities, which are culturally restrictive of change, but by a positive policy initiative to promote multilevel Triple Helix interactions that would help diminish the role of existing vested in interests in social and technological relationships.

The final contribution to this special issue by Matthew Shapiro, Minho So, and Han Woo Park addresses issues relating to the evolution of inter-regional collaboration networks in South Korea, including changing trends in the devolution of innovative activities across the country. Using longitudinal data on scientific co-authorship among Korean researchers located in different regions, the authors show the pattern of the evolving innovation network across the different regions in the country in terms of network centrality, network density, and network fragmentation.

The authors conclude from their empirical analysis that the density of scientific communication flows has deepened in terms of the inter-connectedness of networks, while the centrality of Seoul as the primary research hub has declined, and that network fragmentation is still in evidence to the extent that Seoul has not lost is traditional role as the research broker for the country. It is argued that network fragmentation would decline with the development of Triple Helix relations in the regions which would have the effect of raising Korea's knowledge-based infrastructure, including *inter alia* R&D capabilities. This, however, remains a hypothesis that has yet to be put to the test in the context of South Korea.

The methodology involving head counts of co-authored publications is reflective of the state of networks arising from a process (Triple Helix, national innovation systems or otherwise), but not of the process itself involving the dynamic Triple Helix interactions. Moreover, the robustness of the method would depend on the extent to which co-authored publications have directly or indirectly involved partners from all Triple Helix spheres. Frykfors and Jonsson would, on the other hand, use "cultural analysis" to capture the dynamics in the Triple Helix process including the different social and technological regimes that influence the behaviour of actors in the system. This involves an open search process which allows space for testing some new thoughts and approaches across different phases, including the mapping of key stakeholders; systemic meetings of Triple Helix communities to debate and shape the way forward in terms of selection of "new combinations"; and the development of governance to provide for strategic leadership of innovation and development programmes.

The four articles recognize the need for more work to show how the trilateral relations in the Triple Helix system can be made to function on a sustainable basis, and point out, as Layton-Smith and Bagchi-Sen do, the role of factors like entrepreneurship and skilled labour as "catalysts and integrators within a regional Triple Helix system". These authors note that "entrepreneurs… are neglected within the Triple Helix model which focuses on the system rather than on primary actors;" and that this would not make sense in as much as  a region's ability to innovate is contingent on the individuals' ability to innovate. They attribute this systemic neglect to the Triple Helix model being a top-down based innovation system, with the result that the entrepreneur is at best only implicitly accommodated in the model. Furthermore, the focus of the model on systems ignores the peculiarities of individual actors and treats them as typical players: "…there is no such thing as a typical university and certainly no typical way to become an 'entrepreneurial university'. Both ideas are neglected in Triple Helix model."

In our opinion, the crucial question is which phenomena one wishes to explain in terms of which theories? While each innovation is unique, since emergent and perhaps worthy a thick description, the relevance of case studies for understanding the dynamics of innovation can only be specified if the model is specified in terms of expectations. The observations can inform the expectations and drive us discursively to reformulations. For example, while Nelson & Winter (1977 and 1982) first distinguished between market and non-market selection mechanisms (cf. Von Hippel, 1988), the interactions between and among three selection environments can be expected to generate a complex dynamics. Whereas at a too high level of abstraction all specific data can be made to fit such a complex model, empirical studies force us to be specific, for example, in explaining whether the trajectories can be considered as uniquely constructed advantages which cannot easily be imitated or as instantiations of an emerging abstract regime within which competition is in a different phase. The more sophisticated such a model, the better one may be able to specify what (and why?) one is able to learn from which case studies.